\documentclass[aps,prb,showpacs,reprint,groupedaddress,showkeys,floatfix]{revtex4-1}

\usepackage{graphicx}% Include figure files
\usepackage{dcolumn}% Align table columns on decimal point
\usepackage{bm}% bold math
\usepackage[T1]{fontenc}
\usepackage{amsmath,amssymb}
\usepackage{bm} 
\usepackage{siunitx}
\usepackage{booktabs}
\usepackage{multirow}
\usepackage{lipsum} 
\usepackage{hyperref}% add hypertext capabilities
\begin{document}

% \preprint{APS/123-QED}

\title{Low-temperature infrared dielectric function of hyperbolic $\alpha$-quartz}% Force line breaks with \\

\author{Christopher J. Winta}
\author{Martin Wolf}
\author{Alexander Paarmann}

\email{alexander.paarmann@fhi-berlin.mpg.de}
\affiliation{Department of Physical Chemistry, Fritz-Haber-Institut der Max-Planck-Gesellschaft, Berlin}


\date{\today}% It is always \today, today,

\begin{abstract}
We report the infrared dielectric properties of $\alpha$-quartz in the temperature 		range from \SIrange{1.5}{200}{\kelvin}. Using an infrared free-electron laser, far-infrared reflectivity spectra of a single crystal $y$-cut were acquired along both principal axes, under two different incidence angles, in S- and P-polarization. These experimental data have been fitted globally for each temperature with a multi-oscillator model, allowing to extract frequencies and damping rates of the ordinary and extraordinary, transverse and longitudinal optic phonon modes, and hence the temperature-dependent dispersion of the infrared dielectric function. The results are in line with previous high-temperature studies, allowing for a parametrized description of all temperature-dependent phonon parameters and the resulting dielectric function from \SI{1.5}{\kelvin} up to the $\alpha$-$\beta$-phase transition temperature, $T_C=\SI{846}{\kelvin}$. Using these data, we predict remarkably high quality factors for polaritons in $\alpha$-quartz's hyperbolic spectral region at low temperatures.
\end{abstract}

 %Use showkeys class option if keyword
\maketitle

%\tableofcontents

\section{Introduction}

To date, quartz is of great technological importance, mainly thanks to its pronounced piezoelectricity and abundance in nature.\cite{Bosak2012} Despite being extremely well-studied, quartz and, in particular, its lattice dynamics have been subject to recent research. For instance, its broken inversion symmetry makes it an excellent model system for nonlinear optical techniques, including SFG and SHG phonon spectroscopy.\cite{Liu2008,Winta2018} Often, however, these advanced techniques require good knowledge of the linear optical behavior of the sample which is fully described by its frequency $\omega$-dependent dielectric tensor, $\overset{\text{\tiny$\bm\leftrightarrow$}}{\varepsilon}(\omega)$.\cite{Liu2008,Paarmann2015,Paarmann2016,Winta2018} \par
The dielectric properties of $\alpha$-quartz are of particular interest.\cite{Spitzer1961,Merten1968,Falge1973,Gervais1975,Humlicek1994} The uniaxial crystal naturally features hyperbolicity in the far-infrared (far-IR), i.e., its diagonal permittivity tensor has both, positive and negative principal components, such that $\operatorname{Re}(\varepsilon_\parallel)\cdot\operatorname{Re}(\varepsilon_\perp) < 0$.\cite{Silva2012,Passler2017} This results in the momentum $k$-space's isofrequency surfaces being open hyperboloids (as opposed to, e.g., closed spheres for isotropic media), supporting high-$k$ waves.\cite{Guo2012, Caldwell2014, Dai2015, Li2018, Ma2018} These states, thanks to their large spatial frequency, can be utilized in nanophotonic devices for, e.g., subdiffractional imaging or nanolithography using hyperlenses.\cite{Liu2007, Xiong2009} Moreover, $\alpha$-quartz exhibits both, type I ($\mathrm{Re}(\varepsilon_\parallel)<0$, $\mathrm{Re}(\varepsilon_\perp)>0$) and type II ($\operatorname{Re}(\varepsilon_\parallel)>0$, $\operatorname{Re}(\varepsilon_\perp)<0$) hyperbolicity in close spectral proximity, further adding to its design flexibility.\par
%The dielectric properties of $\alpha$-quartz are of particular interest as they show a promising potential for future device applications due to their natural hyperbolic behavior, opening up opportunities for the development of nanophotonic devices, such as, for example, hyperlenses and waveguides.\cite{Liu2007,Caldwell2014,Dai2015,Li2015} Moreover, quartz features both, Type I ($\mathrm{Re}\ \varepsilon_\parallel < 0$, $\mathrm{Re}\ \varepsilon_\perp > 0$) and Type II ($\mathrm{Re}(\varepsilon_\parallel) > 0$, $\mathrm{Re}(\varepsilon_\perp) < 0$) hyperbolicity, further adding to its design flexibility.\par
Despite its prevailing relevance, low-temperature studies of $\alpha$-quartz's dielectric function have only been reported in the terahertz range.\cite{Davies2018}
In 1975, Gervais and Piriou performed extensive temperature-dependent IR reflectivity measurements on $\alpha$-quartz at high temperatures ranging from \SIrange{295}{975}{\kelvin}.\cite{Gervais1975} Here, we determine $\alpha$-quartz's IR dielectric function at low temperatures, ranging from \SIrange{1.5}{200}{\kelvin}, using an IR free-electron laser (FEL) to acquire reflectivity spectra for various polarization and orientation combinations. These we then fit globally with a multi-oscillator model taking independently into account TO and LO phonon frequencies and damping rates along both principal axes.

%More importantly, quartz's infrared (IR) dielectric properties show a promising potential for future device applications as its natural hyperbolic behavior opens up opportunities for the development of hyperlenses and ...[cite].\par
% utilizing its natural hyperbolic behavior, allowing to ...

%, quartz is subject to numerous spectroscopical studies, including

%At the same time, the infrared (IR) properties of quartz offer promising opportunities for future applications

%Quartz is an abundant mineral that keeps its technologically high relavance to this day, mainly thanks to its pronounced piezoelectricity. However its IR properties are of great interest today

%Albeit being extremely well-studied, quartz keeps its technologically high relevance to this date, mainly thanks to its pronounced piezoelectricity.

% Today, quartz's infrared (IR) properties are of great technologically interest as it is

%However, quartz's infrared (IR) properties present potentially equally interesting opportunities and have been studied

%Quartz is technologically relevant, well-studied, and naturally hyperbolic.\par
%Low temperatures are interesting.\par
%Dielectric properties in the IR are important.

%\section{Reflectivity spectra and analysis}

\section{Experimental setup}
The experimental setup features a noncollinear autocorrelator geometry in reflection with two focused FEL beams impinging on the sample incident at \ang{30} and \ang{60}. The intensity of the reflected beams is then detected by two home-built pyroelectric photodetectors.\par
The sample is placed inside a helium bath cryostat (CryoVac GmbH \& Co KG), facilitating the setup's low-temperature operation, ranging from \SIrange{1.5}{325}{\kelvin} with $<$\SI{0.1}{\kelvin} precision. Optical KRS-5 and diamond windows grant IR accessibility to the sample chamber while the built-in motorized sample stage allows for alignment of the vertical position as well as the polar and azimuthal angle of the sample.\par%In particular the latter allows to align the sample's $c$-axis either vertically or horizontally for selective measurements along the ordinary and extraordinary axes, depending on the polarization of the incident FEL beams.\par
Details on the FEL are reported elsewhere.\cite{Schollkopf2015} %For this experiment, the electron gun was operated at a micropulse repetition rate of \SI{1}{\giga\hertz} with an electron macropulse duration of \SI{10}{\micro\second} and a macropulse repetition rate of \SI{10}{\hertz}. Go on...\par
For these measurements, the electron energy was set to \SI{23}{\mega\electronvolt} and \SI{32}{\mega\electronvolt}, allowing to tune the output wavelength, $\lambda$, from \SIrange{12}{28}{\micro\meter} (\SIrange{350}{850}{\per\centi\meter}) and from \SIrange{7}{18}{\micro\meter} (\SIrange{550}{1400}{\per\centi\meter}), respectively, through variation of the motorized undulator gap. Polarization rotation of the linearly polarized FEL beam is achieved through two subsequent wire-grid polarizers (Thorlabs, Inc.) set to $\ang{45}$ and either $\ang{0}$ or $\ang{90}$ with respect to the incoming P-polarization, for P- and S-polarization, respectively.\par
The sample studied here is an optically polished $\alpha$-quartz $y$-cut single crystal (MaTecK GmbH) with the {[0001]} crystal axis, i.e., the optic $c$-axis, parallel to the surface plane. In this arrangement, both, the ordinary ($E$-type) and extraordinary ($A_2$-type) IR-active vibrational modes can be probed either exclusively or simultaneously, depending on the $c$-axis orientation, adjustable via the sample's azimuthal angle, and the FEL beam polarization.

\section{Dispersion model}
In order to model the IR reflectivity of the $\alpha$-quartz $y$-cut for arbitrary angles of incidence, $\alpha_\mathrm{i}$, as well as horizontal ($c\parallel x$) and vertical ($c\parallel y$) $c$-axis orientations, we evaluate the elements of the Fresnel reflection tensor. These are straightforwardly derived from Maxwell's equations under the boundary condition of in-plane field conservation at the interface with air and read:
%\begin{align}
%\label{eq:R_xx} R_{xx} &= -\frac{\varepsilon_{\perp} k_{\mathrm{i}_z} - k_{\mathrm{t}_z}^\mathrm{e}}{\varepsilon_{\perp} k_{\mathrm{i}_z} + k_{\mathrm{t}_z}^\mathrm{e}},\\
%\label{eq:R_yy} R_{yy} &= \frac{k_{\mathrm{i}_z} - k_{\mathrm{t}_z}^\mathrm{o}}{k_{\mathrm{i}_z} + k_{\mathrm{t}_z}^\mathrm{o}},\\
%\label{eq:R_zz} R_{xx} &= \frac{\varepsilon_{\perp} k_{\mathrm{i}_z} - k_{\mathrm{t}_z}^\mathrm{e}}{\varepsilon_{\perp} k_{\mathrm{i}_z} + k_{\mathrm{t}_z}^\mathrm{e}}.
%\end{align}

\begin{align}
\label{eq:R_xx} R_{xx} &= 
\begin{cases}
-\frac{\varepsilon_{\parallel} k_{\mathrm{i}_z} - k_{\mathrm{t}_z}^\mathrm{e}}{\varepsilon_{\parallel} k_{\mathrm{i}_z} + k_{\mathrm{t}_z}^\mathrm{e}} &\text{for $c\parallel x$,}\\
-\frac{\varepsilon_{\perp} k_{\mathrm{i}_z} - k_{\mathrm{t}_z}^\mathrm{o}}{\varepsilon_{\perp} k_{\mathrm{i}_z} + k_{\mathrm{t}_z}^\mathrm{o}} &\text{for $c\parallel y$,}
\end{cases}\\
\label{eq:R_yy} R_{yy} &= 
\begin{cases}
\frac{k_{\mathrm{i}_z} - k_{\mathrm{t}_z}^\mathrm{o}}{k_{\mathrm{i}_z} + k_{\mathrm{t}_z}^\mathrm{o}} &\text{for $c\parallel x$,}\\
\frac{k_{\mathrm{i}_z} - k_{\mathrm{t}_z}^\mathrm{e}}{k_{\mathrm{i}_z} + k_{\mathrm{t}_z}^\mathrm{e}} &\text{for $c\parallel y$,}
\end{cases}\\
\label{eq:R_zz} R_{zz} &= 
\begin{cases}
\frac{\varepsilon_{\parallel} k_{\mathrm{i}_z} - k_{\mathrm{t}_z}^\mathrm{e}}{\varepsilon_{\parallel} k_{\mathrm{i}_z} + k_{\mathrm{t}_z}^\mathrm{e}} &\text{for $c\parallel x$,}\\
\frac{\varepsilon_{\perp} k_{\mathrm{i}_z} - k_{\mathrm{t}_z}^\mathrm{o}}{\varepsilon_{\perp} k_{\mathrm{i}_z} + k_{\mathrm{t}_z}^\mathrm{o}} &\text{for $c\parallel y$.}
\end{cases}
\end{align}
Here, $k_{\mathrm{i}_z}$, $k_{\mathrm{t}_z}^\mathrm{o}$, and $k_{\mathrm{t}_z}^\mathrm{e}$ denote, respectively, the normal-to-surface ($z$) components of the complex wave vectors of the incoming, and ordinary and extraordinary transmitted waves inside the sample. They read:\cite{Mosteller1968}
\begin{align}
\label{eq:k_i} k_{\mathrm{i}_z} &= 2\pi\omega\cos\alpha_\mathrm{i},\\
\label{eq:k_o} k_{\mathrm{t}_z}^\mathrm{o} &= 2\pi\omega\sqrt{\varepsilon_\perp-\sin^2\alpha_\mathrm{i}},\\
\label{eq:k_e} k_{\mathrm{t}_z}^\mathrm{e} &=
\begin{cases}
2\pi\omega\sqrt{\varepsilon_\parallel-\frac{\varepsilon_\parallel}{\varepsilon_\perp}\sin^2\alpha_\mathrm{i}} &\text{for $c\parallel x$,}\\
2\pi\omega\sqrt{\varepsilon_\parallel-\sin^2\alpha_\mathrm{i}} &\text{for $c\parallel y$.}
\end{cases}
\end{align}
All of these dispersive quantities are ultimately dependent on the elements of the diagonal dielectric tensor, $\varepsilon_\perp$ and $\varepsilon_\parallel$, perpendicular and parallel to the optic $c$-axis, respectively. The highly dispersive dielectric functions, $\varepsilon_\perp(\omega)$ and $\varepsilon_\parallel(\omega)$, can be described using a four-parameter semi-quantum (FPSQ) model:\cite{Adachi1999,Gervais1974,Gervais1975}
%In order to describe $\varepsilon_\perp(\omega)$ and $\varepsilon_\parallel(\omega)$, we use a multi-oscillator model in its product form:\cite{Adachi1999,Gervais1975}

%where $\varepsilon_x$ denotes the in-plane dielectric function. $k_{\mathrm{i}_z} = 2\pi\omega\cos\alpha_\mathrm{i}$ is the $z$-component of the of the complex wave vector of the incoming waves, while $k_{\mathrm{t}_z}^\mathrm{o} = 2\pi\omega\sqrt{\varepsilon_x-\frac{\varepsilon_x}{\varepsilon_z}\sin^2\alpha_\mathrm{i}}$ and $k_{\mathrm{t}_z}^\mathrm{e} = 2\pi\omega\sqrt{\varepsilon_x\sin^2\alpha_\mathrm{i}}$ denote the $z$-components of the complex wave vectors of the ordinary and extraordinary transmitted waves inside the sample, respectively.\cite{Mosteller1968}\par
%All of the dispersive quantities mentioned above are ultimately dependent on the dielectric function $\varepsilon(\omega)$ which we model using a multioscillator model in its product form:\cite{Gervais1975}

\begin{equation} \label{eq:multioscillator}
\varepsilon(\omega) = \varepsilon_\infty \prod_j \frac{\Omega_{\mathrm{LO}_j}^2 - \omega^2 - \mathrm{i}\gamma_{\mathrm{LO}_j}\omega}{\Omega_{\mathrm{TO}_j}^2 - \omega^2 - \mathrm{i}\gamma_{\mathrm{TO}_j}\omega},
\end{equation}
where $\varepsilon_\infty$ is the high-frequency contribution to the dielectric function. $\Omega_{\mathrm{TO(LO)}_j}$ denotes the TO (LO) phonon frequency of the $j$th vibrational mode, and $\gamma_{\mathrm{TO(LO)}_j}$ its respective damping rate. %We note that the multi-oscillator product form can cause negative imaginary parts of the dielectric function. As this unphysical result only occurs where $\operatorname{Im}(\varepsilon)\approx 0$, it can be avoided with little error by taking the absolute, $\left|\operatorname{Im}(\varepsilon)\right|$, instead, when calculating reflectivities.
We note that the FPSQ model in Eq.~\ref{eq:multioscillator} can result in the imaginary part of the dielectric function, $\operatorname{Im}(\varepsilon)$, taking on negative values for large TO-LO splittings of the damping rates, $\Delta\gamma_j = \gamma_{\mathrm{LO}_j}-\gamma_{\mathrm{TO}_j}$.\cite{Gervais1974} To avoid this unphysical regime of the multi-oscillator model, we later apply a penalty to negative values of $\operatorname{Im}(\varepsilon)$ during the least-squares fitting routine.
\par
Finally, the reflected light intensities for P- and S-polarization, $I_\mathrm{P}$ and $I_\mathrm{S}$, respectively, are then straightforwardly given by:
\begin{align}
I_\mathrm{P} &= \left| R_{xx}E_{\mathrm{i}}\cos\alpha_\mathrm{i}\right|^2 + \left| R_{zz}E_{\mathrm{i}}\sin\alpha_\mathrm{i} \right|^2,\label{eq:int_p}\\
I_\mathrm{S} &= \left| R_{yy}E_{\mathrm{i}}\right|^2, \label{eq:int_s}
\end{align}
%This leaves the TO and LO phonon frequencies, $\Omega_{\mathrm{TO}_j}$ and $\Omega_{\mathrm{LO}_j}$, their respective damping rates, $\gamma_{\mathrm{TO}_j}$ and $\gamma_{\mathrm{LO}_j}$, as well as $\varepsilon_\infty$ as independent variables in the model.
where $E_{\mathrm{i}}$ denotes the incident electric field. This leaves the TO and LO phonon frequencies, $\Omega_{\mathrm{TO}_j}$ and $\Omega_{\mathrm{LO}_j}$, their respective damping rates, $\gamma_{\mathrm{TO}_j}$ and $\gamma_{\mathrm{LO}_j}$, as well as the high-frequency contributions, $\varepsilon_{\infty_\perp}$ and $\varepsilon_{\infty_\parallel}$, as the only independent variables in the model.

\section{Results}
\begin{figure*}
	\centering
	\includegraphics[width=0.9\linewidth]{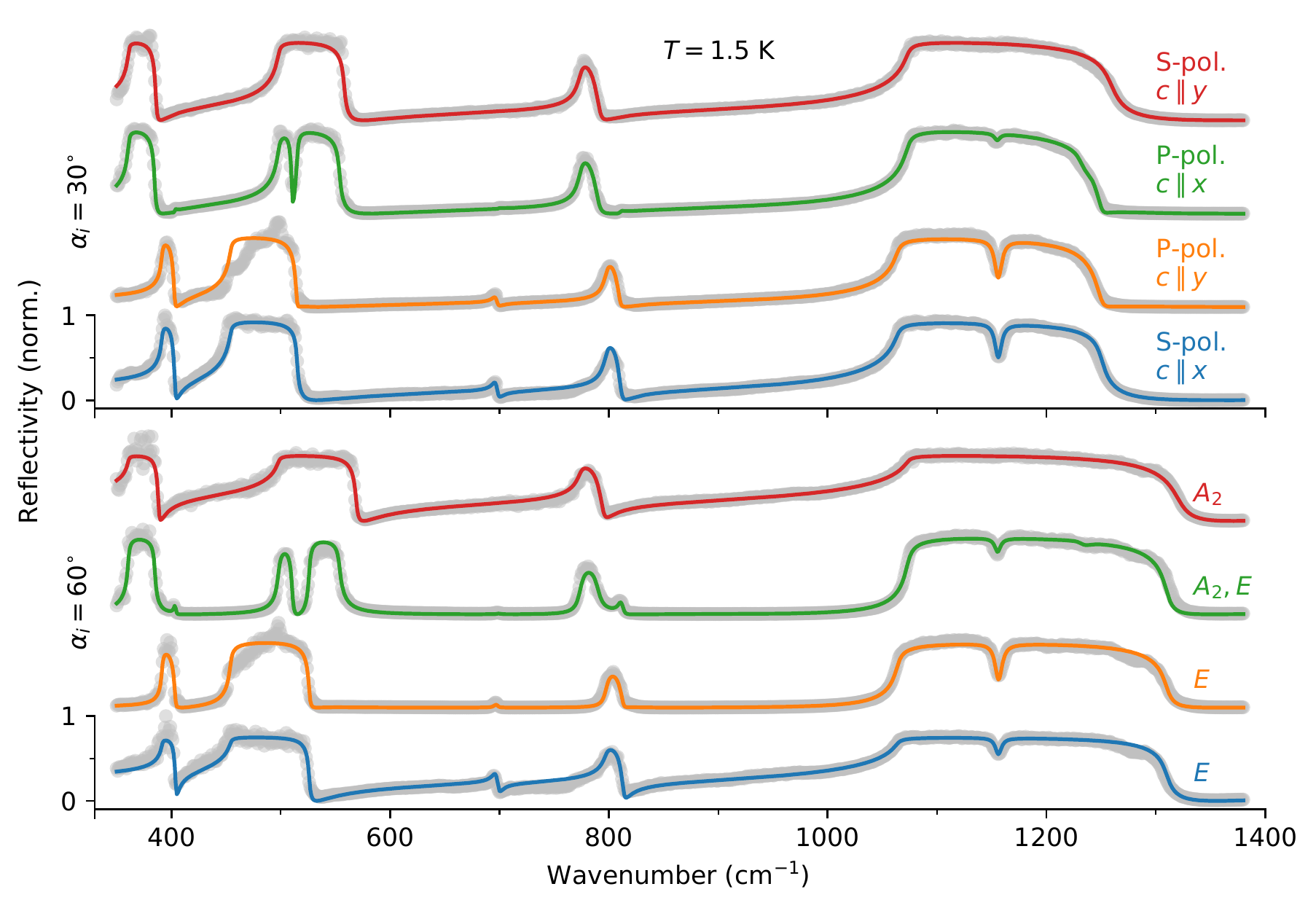}
	\caption{Experimental reflectivity spectra (gray data points) for all FEL beam incidence angles and polarizations as well as sample orientations at the base temperature $T=\SI{1.5}{\kelvin}$ with model fits (solid lines). Depending on the experimental configuration, the measurements are either exclusively sensitive to the ordinary $E$-type modes (blue and orange curves), the extraordinary $A_2$-type modes (red curves), or both (green curves).}
	\label{fig:spectra}
\end{figure*}

%To maximize the amount of data and therefore the fit accuracy,
We have measured all possible combinations of the sample's $c$-axis orientation (vertical and horizontal), the FEL beam polarization (P and S), as well as the angle of incidence ($\ang{30}$ and $\ang{60}$) for four temperatures: \SI{1.5}{\kelvin}, \SI{20}{\kelvin}, \SI{100}{\kelvin}, and \SI{200}{\kelvin}. As an example, we show the experimental data for all geometries and both incidence angles at the base temperature, $T=\SI{1.5}{\kelvin}$, in Fig.~\ref{fig:spectra}.\par

Spectral regions of near-perfect reflectivity emerge between corresponding TO and LO phonon resonances where $\operatorname{Re}(\varepsilon)$ takes on negative values. These so-called Reststrahlen bands are particularly pronounced for modes with large TO-LO splittings, i.e., high oscillator strengths.\cite{Adachi1999} Each measurement configuration, i.e., P- or S-polarization with horizontal or vertical $c$-axis orientation, results in a specific direction of the IR electric field with respect to the principal crystal axes, and thus probes one of the two unique elements of the dielectric tensor, $\varepsilon_\perp$ and $\varepsilon_\parallel$, exclusively, or both simultaneously. For instance, S-polarized reflectivity with $c\parallel x$ as well as P-polarized reflectivity with $c\parallel y$ (blue and orange curves in Fig.~\ref{fig:spectra}, respectively) are exclusively sensitive to the ordinary $E$-type modes and therefore solely probe $\varepsilon_\perp$. Similarly, S-polarized reflectivity with $c\parallel y$ (red curve), probes the extraordinary $A_2$-type modes, thus $\varepsilon_\parallel$. Comparing these spectra of exclusive sensitivity to $\varepsilon_\perp$ and $\varepsilon_\parallel$, e.g. blue vs. red curves in Fig.~\ref{fig:spectra}, respectively, reveals $\alpha$-quartz's pronounced uniaxial anisotropy, being the result of different numbers of IR-active modes and significant frequency shifts between its principal axes. P-polarized reflectivity with $c\parallel x$ (green curve), on the other hand, is sensitive to both principal components. This becomes apparent in the reflectivity spectra (green curves) as features attributable to both, $E$- and $A_2$-type modes, are present, the former being more pronounced for the more grazing \ang{60} incidence angle, i.e., for a larger incoming out-of-plane field component.\par
%through the presence of features caused by the ordinary $E$-type modes whose influence is stronger for the more grazing \ang{60} incidence angle, i.e., larger incoming $z$-field component.
%The sole sensitivity to the ordinary and extraordinary axes can be seen Shifted Reststrahlen edges\par
The experimental spectra have been fitted with a nonlinear least squares regression procedure applied globally to the entire data set for each temperature using Eqs.~{\ref{eq:R_xx}--\ref{eq:int_s}}, yielding the frequencies and damping rates of both, $E$- and $A_2$-type phonon modes, as well as the high-frequency contributions, $\varepsilon_{\infty_\perp}$ and $\varepsilon_{\infty_\parallel}$.\par
The fits are in good agreement with the experimental data across the whole data set. Noticeable deviations from the experimental data can be observed in the lower frequency region from \SIrange{450}{550}{\per\centi\meter} which we suspect are caused by two parasitic spectral contributions: (i) The cryostat windows---although being wedged---cause a Fabry-Perot-like spectral modulation on top of the measured raw data. This modulation happens to be particularly pronounced at the lower Reststrahlen region at $\alpha_\mathrm{i}=\ang{60}$ and becomes more prominent where signal levels are constantly high, i.e., in Restrahlen bands. (ii) CO$_2$ bending modes in air cause strong IR absorption in this spectral region which we minimize by flooding the optical setup with N$_2$ gas. Both of these unwanted effects are being corrected for by measuring reference spectra in S- and P-polarization using a gold mirror as a nominally perfect reflector in place of the sample and dividing the raw reflectivity spectra by their corresponding reference. However, a certain spectral modulation remains and is observable in the corrected spectra. Nevertheless, these modulations do not have a strong influence on the fit procedure as positions and widths of Reststrahlen edges remain largely unaffected.\par
The resulting fit parameters from all measurements for $E$- and $A_2$-type phonon modes are summarized in Tables~\ref{tab:e_modes} and \ref{tab:a2_modes}, respectively. Here, damping rates that fall below the FEL linewidth of $\Delta\omega\gtrsim \SI{2}{\per\centi\meter}$ are indicated as <2.0. Values for $\varepsilon_{\infty_\perp}$ and $\varepsilon_{\infty_\parallel}$ are averaged over all four measured temperatures as the values do not show any significant temperature dependence.

\begin{table}
\centering
\footnotesize
\caption{Results for $E$-type phonon mode frequencies and damping rates used as free parameters to fit the experimental reflectivity spectra. Values for $\varepsilon_{\infty_\perp}$ have been averaged over all temperatures.}
\begin{tabular}{@{}crSS[table-number-alignment=center,table-format=<1.2]SS[table-number-alignment=center,table-format=<1.2]@{}}
\toprule
\multirow{2}{*}{$j$} & \multicolumn{1}{c}{$T$} & \multicolumn{1}{c}{$\Omega_{\mathrm{TO}_j}$} & \multicolumn{1}{c}{$\gamma_{\mathrm{TO}_j}$} & \multicolumn{1}{c}{$\Omega_{\mathrm{LO}_j}$} & \multicolumn{1}{c}{$\gamma_{\mathrm{LO}_j}$} \\
  & \multicolumn{1}{c}{(\si{\kelvin})} & \multicolumn{1}{c}{(\si{\per\centi\meter})} & \multicolumn{1}{c}{(\si{\per\centi\meter})} & \multicolumn{1}{c}{(\si{\per\centi\meter})} & \multicolumn{1}{c}{(\si{\per\centi\meter})} \\
\midrule
3 & 1.5 & 391.5 & <2.0& 403.0 & <2.0  \\ % 0.7 1.3
  & 20  & 392.1 & <2.0& 402.8 & <2.0 \\ % 1.0 1.4
  & 100 & 391.9 & <2.0& 403.0 & <2.0 \\ % 0.5 1.3
  & 200 & 392.4 & <2.0& 402.8 & 2.6 \\[3pt] % 1.5 2.1
4 & 1.5 & 454.0 & 2.6& 510.5 & <2.0 \\ % 1.8 1.7
  & 20  & 452.4 & 3.4 & 510.4 & 2.3 \\
  & 100 & 453.6 & 2.2 & 510.0 & 2.1 \\ % 1.6 2.4
  & 200 & 451.2 & 4.0 & 508.8 & 2.8 \\[3pt]
5 & 1.5 & 695.9 & 4.9 & 698.4 & 4.0 \\
  & 20  & 695.8 & 4.9 & 698.5 & 4.5 \\
  & 100 & 696.0 & 4.2 & 698.1 & 3.9 \\
  & 200 & 695.3 & 5.6 & 697.7 & 5.0 \\[3pt]
6 & 1.5 & 797.2 & 4.8 & 810.0 & 4.3 \\
  & 20  & 797.2 & 4.4 & 810.2 & 5.0 \\
  & 100 & 796.9 & 5.1 & 809.9 & 4.1 \\
  & 200 & 796.5 & 6.0 & 809.1 & 5.2 \\[3pt]
7 & 1.5 & 1063.7 & 6.1 & 1230.7 & 8.2 \\
  & 20  & 1062.9 & 6.3 & 1231.9 & 10.9 \\
  & 100 & 1063.9 & 6.8 & 1231.2 & 12.0 \\
  & 200 & 1063.0 & 7.1 & 1230.0 & 12.1 \\[3pt]
8 & 1.5 & 1157.2 & 6.2 & 1154.9 & 6.1 \\
  & 20  & 1156.9 & 6.9 & 1154.4 & 6.2 \\
  & 100 & 1157.0 & 6.2 & 1154.9 & 6.0 \\
  & 200 & 1157.0 & 7.2 & 1154.8 & 6.3 \\[3pt]
  & & \multicolumn{2}{l}{$\varepsilon_{\infty_\perp} = 2.296$}\\
\bottomrule
\end{tabular}
\label{tab:e_modes}
\end{table}

\begin{table}
\centering
\footnotesize
\caption{Results for $A_2$-type phonon mode frequencies and damping rates used as free parameters to fit the experimental reflectivity spectra. Values for $\varepsilon_{\infty_\parallel}$ have been averaged over all temperatures.}
\begin{tabular}{@{}crSS[table-number-alignment=center,table-format=<1.2]SS[table-number-alignment=center,table-format=<1.2]@{}}
\toprule
\multirow{2}{*}{$j$} & \multicolumn{1}{c}{$T$} & \multicolumn{1}{c}{$\Omega_{\mathrm{TO}_j}$} & \multicolumn{1}{c}{$\gamma_{\mathrm{TO}_j}$} & \multicolumn{1}{c}{$\Omega_{\mathrm{LO}_j}$} & \multicolumn{1}{c}{$\gamma_{\mathrm{LO}_j}$} \\
  & \multicolumn{1}{c}{(\si{\kelvin})} & \multicolumn{1}{c}{(\si{\per\centi\meter})} & \multicolumn{1}{c}{(\si{\per\centi\meter})} & \multicolumn{1}{c}{(\si{\per\centi\meter})} & \multicolumn{1}{c}{(\si{\per\centi\meter})} \\
\midrule
1 & 1.5 & 360.7 & <2.0 & 384.8 & <2.0 \\ % 0.8 1.1
  & 20  & 360.1 & <2.0 & 384.8 & <2.0 \\ % 0.6 1.0
  & 100 & 360.9 & <2.0 & 384.3 & <2.0 \\ % 1.3 1.3
  & 200 & 361.2 & 2.1 & 385.2 & <2.0 \\[3pt]
2 & 1.5 & 497.9 & 3.1 & 553.6 & 2.8 \\
  & 20  & 498.4 & 3.6 & 554.4 & 3.3 \\
  & 100 & 498.0 & 3.2 & 553.1 & 3.1 \\
  & 200 & 496.8 & 4.7 & 552.2 & 4.0 \\[3pt]
3 & 1.5 & 773.7 & 5.4 & 789.9 & 6.3 \\
  & 20  & 774.4 & 5.6 & 789.9 & 6.5 \\
  & 100 & 774.1 & 5.8 & 789.9 & 7.1 \\
  & 200 & 775.3 & 5.9 & 788.8 & 6.8 \\[3pt]
4 & 1.5 & 1073.0 & 6.2 & 1238.7 & 12.4 \\
  & 20  & 1072.7 & 3.5 & 1241.2 & 11.2 \\
  & 100 & 1072.8 & 4.9 & 1239.2 & 11.1 \\
  & 200 & 1070.9 & 5.3 & 1239.6 & 11.5 \\[3pt]
  & & \multicolumn{2}{l}{$\varepsilon_{\infty_\parallel} = 2.334$}\\
\bottomrule
\end{tabular}
\label{tab:a2_modes}
\end{table}

\section{Discussion}
\begin{figure}
	\centering
	\includegraphics[width=\linewidth]{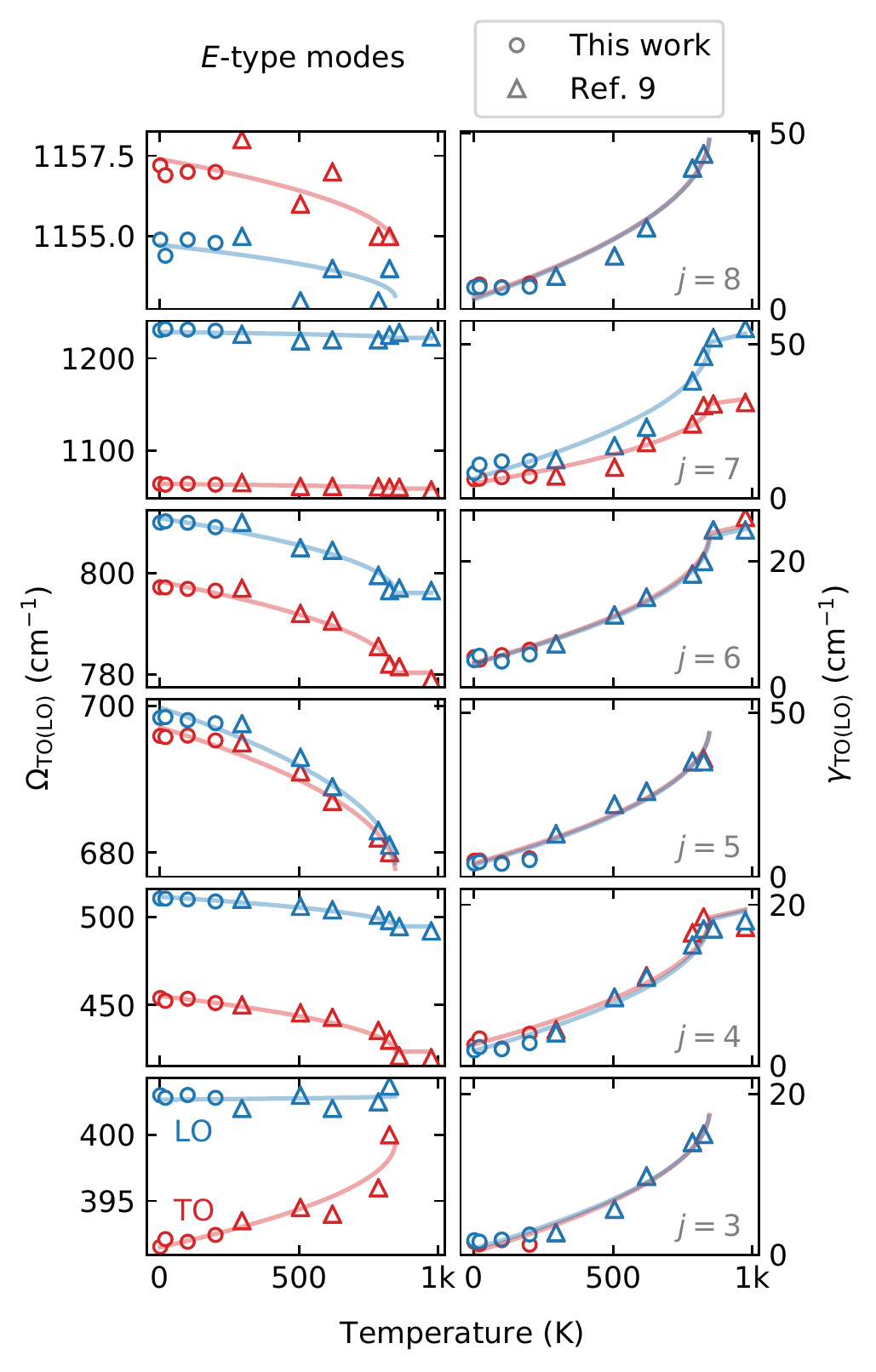}
	\caption{Thermal evolution of phonon frequencies (left side) and damping rates (right side) for the ordinary $E$-type phonon modes. Solid lines indicate curve fits.}
	\label{fig:temp_dep_e}
\end{figure}

\begin{figure}
	\centering
	\includegraphics[width=\linewidth]{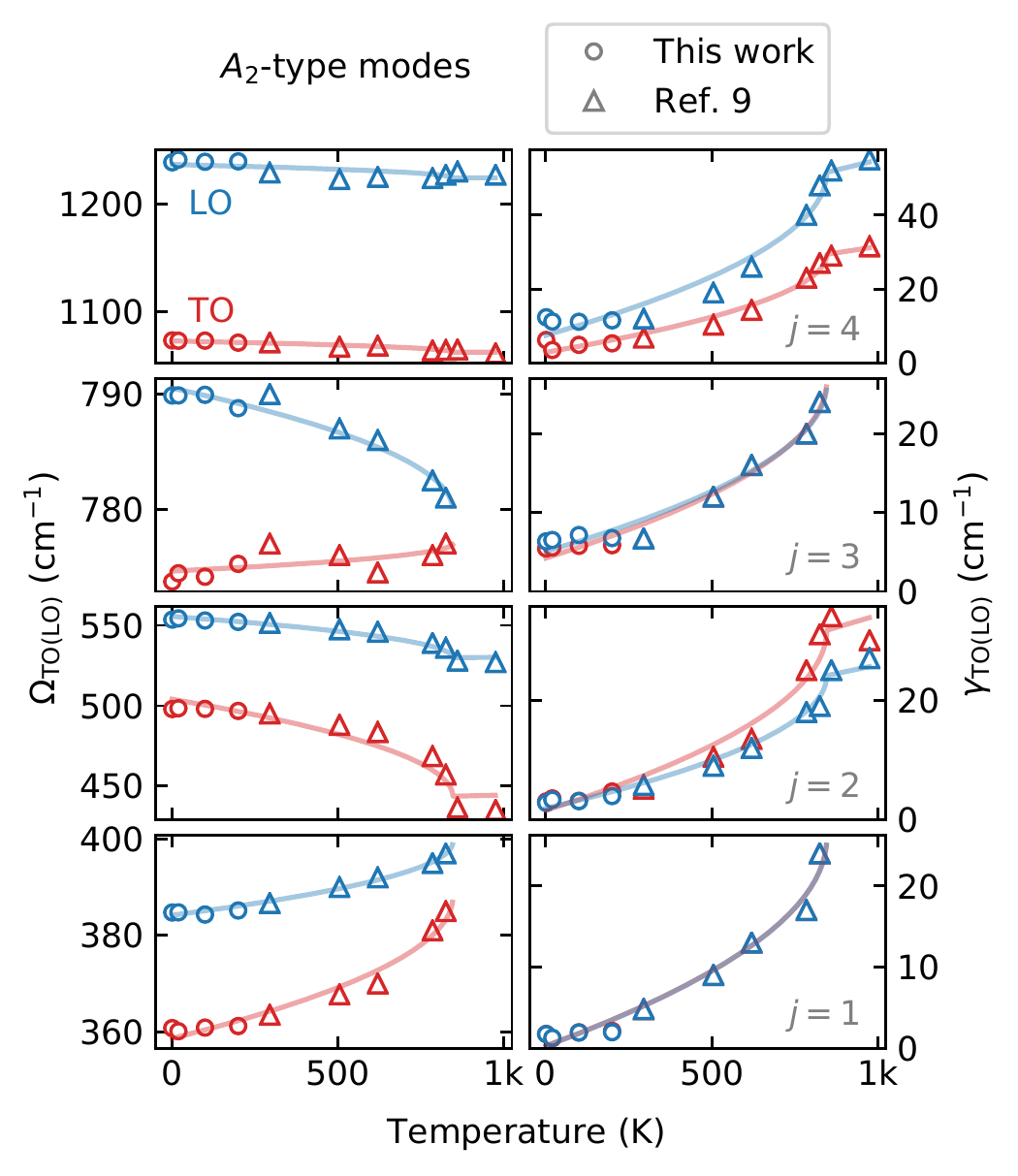}
	\caption{Thermal evolution of phonon frequencies (left side) and damping rates (right side) for the extraordinary $A_2$-type phonon modes. Solid lines indicate curve fits.}
	\label{fig:temp_dep_a2}
\end{figure}

In Figs.~\ref{fig:temp_dep_e} and \ref{fig:temp_dep_a2}, we plot the temperature dependence of our fit results for TO and LO phonon frequencies and damping rates together with the values determined by Gervais and Piriou who studied the dielectric properties of $\alpha$-quartz at high temperatures.\cite{Gervais1975} Overall, our results are largely consistent with Gervais and Piriou's previous work as the extension of our low-temperature results to the high-temperature values taken from Ref.~\citenum{Gervais1975} is rather gradual. Notably, the majority of modes still experience a significant decrease in damping rates below \SI{295}{\kelvin}. In particular, the spectrally lower $E$-type $j=3,4$ and $A_2$-type $j=1,2$ modes which support $\alpha$-quartz's pronounced hyperbolicity, experience a reduced damping rate by nearly a factor of 2 compared to room temperature.\par
To describe the temperature dependence of both, the phonon frequencies and damping rates, we apply a power law fit with the vertex at $T_C$ to the entire temperature range including Gervais and Piriou's high-temperature data:\cite{Polyakova2014}
\begin{equation}
y_j(T) = \left|y_j(\SI{0}{\kelvin}) + k (T_C-T)^{1/2}\right|,
\end{equation}
where $y_j$ denotes either the phonon frequency, $\Omega_j$, or the damping rate, $\gamma_j$, of the $j$th mode. The fitted curves are also shown in Figs.~\ref{fig:temp_dep_e} and \ref{fig:temp_dep_a2} as solid lines which describe the temperature-dependent behavior of $\Omega_j(T)$ and $\gamma_j(T)$ with good accuracy.%We note, however, that the decription for quartz's $\beta$-phase is 
%However, one distinct difference can be observed with the $A_2$-type $j=3$ mode. %First, the $E$-type mode $j=9$ has not been taken into account in Ref.~\citenum{Gervais1975}. In fact, an accurate determination of the frequencies and damping rates for this particularly weak mode is difficult as its signature in the reflectivity spectra is extremely faint. Consequently, measurement errors take on large values as indicated by the shaded areas in Fig~\ref{fig:temp_dep_a2}. 
%While Gervais and Piriou found identical damping rates of the respective TO and LO modes, our fit procedure results in significantly different values for $\gamma_\mathrm{TO}^{j=3}$ and $\gamma_\mathrm{LO}^{j=3}$ (see Fig.~\ref{fig:temp_dep_a2}) with $\gamma_\mathrm{TO}^{j=3}$ being well in line with Gervais' and Piriou's result, but $\gamma_\mathrm{LO}^{j=3}$ being on average $\sim$\SI{4}{\per\centi\meter} higher across the entire temperature range. Apart from that, we observe gradual transitions from our low-temperature data to the high-temperature values taken from Ref.~\citenum{Gervais1975}.

From the phonon frequencies and damping rates, we can now easily compute the dielectric functions, $\varepsilon_\perp(\omega)$ and $\varepsilon_\parallel(\omega)$, using Eq.~\ref{eq:multioscillator}. %A fully parametrized and temperature-dependent dielectric tensor based on the power law fits can be found in the Supplemental Material as a Python Jupyter Notebook and a MATLAB script. 
A fully parametrized and temperature-dependent dielectric tensor based on the power law fits to our as well as Gervais and Piriou's high-temperature data can be calculated using the Python Jupyter Notebook or MATLAB script supplied in Ref.~\onlinecite{Winta2019} which give physical results for $\varepsilon_\perp(\omega)$ and $\varepsilon_\parallel(\omega)$ over the full wavelength range studied (\SIrange{350}{1380}{\per\centi\meter}). We note, that we restrict the validity of the resulting dielectric function to quartz's $\alpha$-phase, where the power law fits describe the data with high accuracy. However, our simple model fails to describe the resonant behavior of damping rates at $T_C$ and leads to instabilities of the FPSQ model for $T>T_C$.\par
\begin{figure*}
	\centering
	\includegraphics[width=\linewidth]{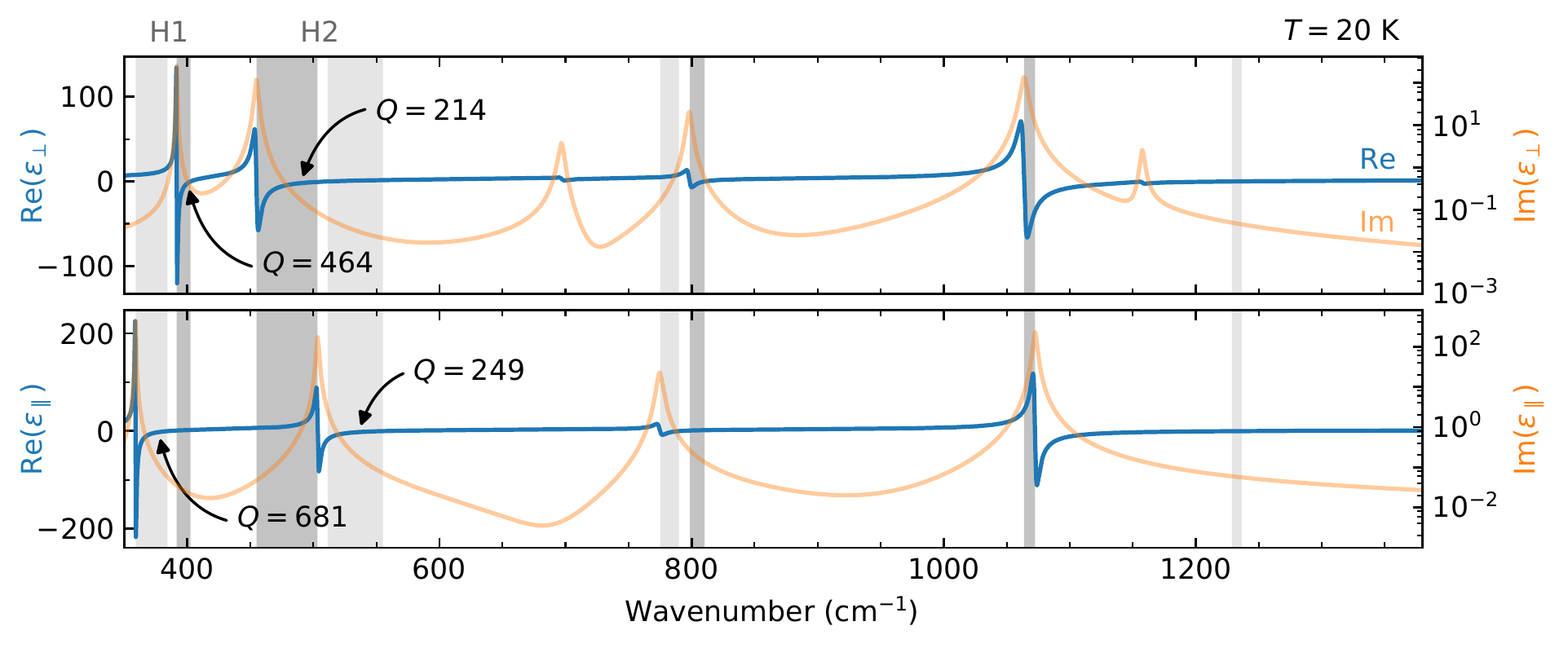}
	\caption{Resulting dielectric functions $\varepsilon_\perp(\omega)$ and $\varepsilon_\parallel(\omega)$ at $T=\SI{20}{\kelvin}$. Left axes (blue) indicate the real part of the dielectric functions, while the right axes (orange) indicate their imaginary parts (on a semilogarithmic scale). Shaded areas mark type I (light grey) and type II (dark grey) hyperbolic bands. H1 and H2 indicate the two pronounced pairs of hyperbolic bands in $\alpha$-quartz's lower spectral region. $Q$-factors are marked where $\operatorname{Re}(\varepsilon)=-2$.}%Grey shades marks the spectral regions where $\operatorname{Re}\left(\varepsilon_\perp\right)$ and $\operatorname{Re}\left(\varepsilon_\parallel\right)$ have opposite signs, i.e., $\alpha$-quartz exhibits hyperbolicity. Black arrows indicate quality factors $Q$ at $\operatorname{Re}(\varepsilon)=-2$.}
	\label{fig:epsilon}
\end{figure*}

We exemplarily plot the resulting real and imaginary parts of $\varepsilon_\perp(\omega)$ and $\varepsilon_\parallel(\omega)$ at $T=\SI{20}{\kelvin}$ (Fig.~\ref{fig:epsilon}), a typical temperature where numerous current polariton studies are being performed, e.g. using cryogenic scanning near-field optical microscopy.\cite{Ni2018} Light grey shades mark type I hyperbolic regions where $\operatorname{Re}(\varepsilon_\parallel)<0$ and $\operatorname{Re}(\varepsilon_\perp)>0$, while darker shades indicate type II hyperbolic bands where $\operatorname{Re}(\varepsilon_\parallel)>0$ and $\operatorname{Re}(\varepsilon_\perp)<0$. Especially in the lower spectral region, between $\sim$\SI{360}{\per\centi\meter} and $\sim$\SI{550}{\per\centi\meter}, pronounced hyperbolic bands of both types, I and II, emerge.

To assess whether these hyperbolic bands are suitable for potential nanophotonic applications, we evaluate the quality factor $Q$ as a figure of merit (FOM) which reads:\cite{Caldwell2015}
%Equipped with the dielectric functions, we can now assess $\alpha$-quartz's potential for nanophotonic applications. For that, we introduce the quality factor $Q$ as a figure of merit (FOM) which reads:\cite{Caldwell2015}
%One common figure of merit is the quality factor $Q$ which quantifies... and can be estimated as:\cite{Caldwell2015}
\begin{equation}
Q=\frac{\omega\frac{\mathrm{d}\operatorname{Re}(\varepsilon)}{\mathrm{d}\omega}}{2\operatorname{Im}(\varepsilon)}\Biggm\lvert_{\operatorname{Re}(\varepsilon) = -2.}
\end{equation}
For comparability, we determine $Q$ in all hyperbolic bands in the lower spectral region between \SI{361}{\per\centi\meter} and \SI{554}{\per\centi\meter} at $\operatorname{Re}(\varepsilon)=-2$. Thus, we here follow the common practice in the field of plasmonics where $\operatorname{Re}(\varepsilon) = -2$ marks the peak of the absorption cross-section of a spherical nanoparticle in air due to a localized surface plasmon resonance.\cite{Caldwell2015} Another important property of hyperbolic materials is the ratio $\varepsilon_\perp/\varepsilon_\parallel$ as it defines the rigid propagation direction of hyperbolic polaritons in the given material and plays a crucial role in the design of, e.g., hyperlenses.\cite{Caldwell2014} For this reason, we also evaluate $\operatorname{Re}(\varepsilon_{\perp(\parallel)})$ at $\operatorname{Re}(\varepsilon_{\parallel(\perp)})=-2$. Both FOMs are given in Table~\ref{tab:FOM} for $\alpha$-quartz (SiO$_2$) at \SI{20}{\kelvin} and \SI{300}{\kelvin} as well as hexagonal boron nitride (h-BN) for comparison. Hexagonal boron nitride has recently been subject to various studies, making use of its natural hyperbolicity and excellent $Q$-factors.\cite{Caldwell2014,Caldwell2015,Giles2018,Dai2014,Dai2015a,Li2018} Here, we refer to h-BN as a benchmark system to evaluate $\alpha$-quartz's potential for nanophotonic device applications.\par

\begin{table*}
\centering
\footnotesize
\caption{Comparison of $\alpha$-quartz's hyperbolic regions, H1 and H2, at \SI{20}{\kelvin} and \SI{300}{\kelvin}, with the naturally hyperbolic hexagonal boron nitride, comprising the parametrized TO and LO phonon frequencies, $Q$-factors at $\operatorname{Re}(\varepsilon)=-2$, and $\operatorname{Re}(\varepsilon_{\perp(\parallel)})$ where $\operatorname{Re}(\varepsilon_{\parallel(\perp)})=-2$ for both principal crystal axes, based on the power law fits.}
\begin{tabular}{@{}rSSSSSSSSS@{}}
\toprule
\multirow{2}{*}{Material} & \multicolumn{1}{c}{$T$ (\si{\kelvin})} & \multicolumn{2}{c}{$\Omega_\perp$ (\si{\per\centi\meter})} & \multicolumn{2}{c}{$\Omega_\parallel$ (\si{\per\centi\meter})} & \multicolumn{2}{c}{$Q\vert_{\operatorname{Re}(\varepsilon)=-2}$} & \multicolumn{2}{c}{$\operatorname{Re}(\varepsilon_{\perp(\parallel)})\vert_{\operatorname{Re}(\varepsilon_{\parallel(\perp)})=-2}$}\\
%  \multicolumn{1}{c}{$\operatorname{Re}(\varepsilon_{\perp(\parallel)})$ at $\operatorname{Re}(\varepsilon_{\parallel(\perp)})=-2$}
 %& \multicolumn{1}{c}{(\si{\kelvin})} & \multicolumn{2}{c}{(\si{\per\centi\meter})} & \multicolumn{2}{c}{(\si{\per\centi\meter})} & \multicolumn{1}{c}{}\\
\cmidrule(r){3-4}\cmidrule(r){5-6}\cmidrule(r){7-8}\cmidrule(r){9-10}
  & & \multicolumn{1}{c}{TO} & \multicolumn{1}{c}{LO} & \multicolumn{1}{c}{TO} & \multicolumn{1}{c}{LO} & \multicolumn{1}{c}{$Q_\perp$} & \multicolumn{1}{c}{$Q_\parallel$} & \multicolumn{1}{c}{$\operatorname{Re}(\varepsilon_\perp)$} & \multicolumn{1}{c}{$\operatorname{Re}(\varepsilon_\parallel)$}\\
\midrule
$\alpha$-SiO$_2$ H1 & 20 & 391.6 & 402.7 & 358.9 & 384.4 & 464 & 681 & 11.0 & 2.2 \\
\text{[}\onlinecite{Gervais1975}\text{]} & 300 & 393.1 & 402.7 & 364.3 & 387.2 & 87 & 69 & 11.9 & 2.2\\[3pt]
$\alpha$-SiO$_2$ H2 & 20 & 454.8 & 511.3 & 503.4 & 555.3 & 214 & 249 & 1.1 & 15.0 \\
\text{[}\onlinecite{Gervais1975}\text{]} & 300 & 449.0 & 508.2 & 492.2 & 550.6 & 91 & 76 & 0.9 & 23.9 \\[3pt]
h-BN [\onlinecite{Caldwell2014}]  & 300 & 1360 & 1614 & 760 & 825 & 221 & 399 & 8.0 & 2.8 \\
\bottomrule
\end{tabular}
\label{tab:FOM}
\end{table*}

The comparison shows that $\alpha$-quartz at low temperatures offers very good $Q$-factors, even surpassing those reported for h-BN at room temperature. We note, that reflectivity-based methods can only offer limited sensitivity to the small off-resonance imaginary part of the dielectric function. Therefore, $Q$-factors should be understood as estimates and a precise determination of the polariton performance requires a more direct measurement. The analysis also shows that $\alpha$-quartz enables the high-$k$ states characteristic for hyperbolic materials in a distinctly different spectral range, i.e., between \SI{361}{\per\centi\meter} and \SI{554}{\per\centi\meter}, as opposed to h-BN which exhibits type I hyperbolicity from \SI{760}{\per\centi\meter} to \SI{825}{\per\centi\meter} and type II hyperbolicity from \SI{1360}{\per\centi\meter} to \SI{1614}{\per\centi\meter}.
%The comparison shows that $\alpha$-quartz at low temperatures offers very good $Q$-factors, even surpassing those reported for h-BN (at room temperature) while enabling the high-$k$ states characteristic for hyperbolic materials in a distinctly different spectral range, i.e., between \SI{361}{\per\centi\meter} and \SI{554}{\per\centi\meter} as opposed to h-BN which exhibits type I hyperbolicity from \SI{760}{\per\centi\meter} to \SI{825}{\per\centi\meter} and type II hyperbolicity from \SI{1360}{\per\centi\meter} to \SI{1614}{\per\centi\meter}. We note, however, that reflectivity-based methods can only offer limited sensitivity to the small off-resonance imaginary part of the dielectric function. Therefore, $Q$-factors should be understood as estimates and a determination of the real polariton performance requires a more direct measurement.
%in order to determine the real polariton performance, direct measurements are 
%though, that our extracted $Q$-factors respond quite sensitively to the details of the multioscillator parameters and should be undertood as estimates. 
Furthermore, $\alpha$-quartz possesses both, type I and type II hyperbolic bands within the same spectral region, opening up additional design opportunities for nanophotonic devices. Also in terms of propagation direction of hyperbolic polaritons which is defined by the ratio $\varepsilon_\perp/\varepsilon_\parallel$, $\alpha$-quartz offers additional flexibility as its lower pair of hyperbolic bands (indicated as ``H1'' in Fig.~\ref{fig:epsilon} and Table~\ref{tab:FOM}) exhibits similar properties as h-BN whereas band ``H2'' provides significantly different values while being in very close spectral proximity to H1. While the latter is also true at room temperature, $Q$-factors of hyperbolic polaritons experience a substantial improvement by at least a factor of 2 as compared to room temperature, promising superior performance of nanophotonic devices utilizing $\alpha$-quartz's hyperbolicity at cryogenic temperatures.

\section{Conclusion}
We report the dielectric properties of $\alpha$-quartz in the temperature range from \SIrange{1.5}{200}{\kelvin} using IR reflectivity measurements and a global fit procedure to extract the temperature-dependent phonon frequencies and damping rates. From the results, we calculate the in-plane and out-of-plane dielectric function, $\varepsilon_\perp$ and $\varepsilon_\parallel$, respectively, which indicate remarkably high $Q$-factors for polaritons in $\alpha$-quartz's naturally hyperbolic spectral region while offering additional design opportunities over established hyperbolic materials like h-BN.

\subsection*{Acknowledgment}
The authors thank Wieland Sch{\"o}llkopf and Sandy Gewinner for their assistance with operating the free-electron laser.

\end{document}